\documentclass[conference]{IEEEtran}
\IEEEoverridecommandlockouts
\usepackage{cite}
\usepackage{amsmath,amssymb,amsfonts}
\usepackage{algorithmic}
\usepackage{graphicx}
\usepackage{textcomp}
\usepackage{multirow}
\usepackage{tabularx}
\usepackage{lscape}
\usepackage{diagbox}
\usepackage{booktabs}
\usepackage{xcolor}
\usepackage{url}
\def\BibTeX{{\rm B\kern-.05em{\sc i\kern-.025em b}\kern-.08em
    T\kern-.1667em\lower.7ex\hbox{E}\kern-.125emX}}
\begin{document}

\title{BiTSA: Leveraging Time Series Foundation Model for Building Energy Analytics\\
}


\author{
    \IEEEauthorblockN{
    Xiachong Lin\IEEEauthorrefmark{1}, 
    Arian Prabowo\IEEEauthorrefmark{1}, 
    Imran Razzak\IEEEauthorrefmark{1}, 
    Hao Xue\IEEEauthorrefmark{1}, 
    Matthew Amos\IEEEauthorrefmark{2},
    Sam Behrens\IEEEauthorrefmark{2},
    Flora D. Salim\IEEEauthorrefmark{1}} 
    \IEEEauthorblockA{\IEEEauthorrefmark{1}University of New South Wales, Sydney, Australia} 
    \IEEEauthorblockA{\IEEEauthorrefmark{2}CSIRO Energy Centre, Newcastle, Australia} 
    Email: \{dawn.lin, arian.prabowo, imran.razzak, hao.xue1, flora.salim\}@unsw.edu.au \\
    \{matt.amos, sam.behrens\}@csiro.au
}

\maketitle

\begin{abstract}

Incorporating AI technologies into digital infrastructure offers transformative potential for energy management, particularly in enhancing energy efficiency and supporting net-zero objectives. However, the complexity of IoT-generated datasets often poses a significant challenge, hindering the translation of research insights into practical, real-world applications. This paper presents the design of an interactive visualization tool, BiTSA.
The tool enables building managers to interpret complex energy data quickly and take immediate, data-driven actions based on real-time insights. By integrating advanced forecasting models with an intuitive visual interface, our solution facilitates proactive decision-making, optimizes energy consumption, and promotes sustainable building management practices.
BiTSA will empower building managers to optimize energy consumption, control demand-side energy usage, and achieve sustainability goals. The demo video\footnote{\url{https://github.com/Dawnlxc/BiTSA_video}} is made publicly available. 

\end{abstract}


\section{Introduction}
Buildings play a pivotal role in global energy consumption and sustainability efforts, accounting for a significant share of worldwide energy demand. As energy consumption continues to rise, particularly in urban environments, optimizing energy use in buildings has become crucial for achieving sustainability goals and addressing climate change. By focusing on key areas such as energy efficiency, renewable energy integration, smart technologies, and low-carbon solutions, buildings can be transformed into vital contributors to a sustainable future.

The advent of smart technologies and the Internet of Things (IoT) has revolutionized energy management in urban buildings. These technologies enable real-time monitoring and cloud-based data analysis of energy consumption patterns, allowing for advanced strategies like predictive maintenance and demand-response systems~\cite{prabowo2024bts, luo2022three, IEA_EBC_Annex81}. By reducing energy usage during peak demand periods and improving overall efficiency, smart buildings can contribute to greater grid stability and cost reduction. 

However, the data generated by IoT systems in buildings—ranging from energy usage metrics to environmental and comfort parameters—often involves complex dependencies. Without advanced analytical tools, this data is difficult to interpret, especially for non-expert users, creating a barrier to its practical use in everyday building management~\cite{lin2024gap}.

Moreover, effective energy management is essential for reducing operational costs and achieving sustainability targets, but identifying inefficiencies and making timely adjustments remains a significant challenge. While advanced research in energy forecasting provides valuable insights, these models often fail to transition smoothly into real-world applications due to their complexity. Building managers require actionable insights that are easy to implement, yet research models are difficult for day-to-day operational use. To address these challenges, practical and user-friendly tools that can translate the complex analysis of IoT data into actionable insights are urgently needed. Such tools are essential for empowering stakeholders such as building managers to optimize energy consumption, control demand-side energy usage, and achieve sustainability goals. By deploying cutting-edge technologies for building activity monitoring, energy control, and strategy optimization, these tools can enable efficient building management practices, bridging the gap between research and real-world applications.

\begin{figure}[!htb]
    \centering
    \includegraphics[width=\linewidth]{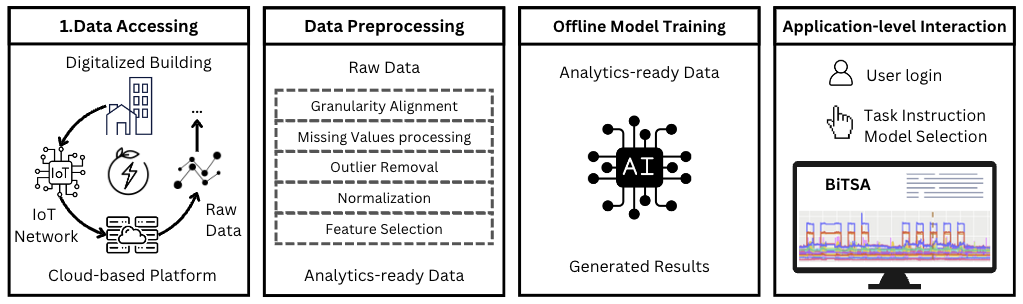}
    \caption{The general framework of BiTSA for assisting building managers in fetching accurate building activities based on pre-trained AI models.}
    \label{fig:pipeline}
\end{figure}

\textbf{B}u\textbf{i}lding\textbf{T}ime\textbf{S}eries\textbf{A}nalytics (BiTSA) will provide building managers with a powerful solution for optimizing energy management, as Figure~\ref{fig:pipeline}. This tool will deliver the following major contributions:
\begin{itemize}
    \item Provides an interactive visualization platform, enabling managers to interpret complex IoT-generated infrastructure data efficiently.
    \item Utilizes pre-trained time-series models to analyze pre-processed data, delivering insights based on historical trends to assist in future energy planning.
    \item Bridges the gap between academic research and practical industry applications by integrating cutting-edge AI technologies, making advanced energy forecasting accessible to building managers and facility operators.
\end{itemize}
The remainder of the paper is organized as follows: Section \ref{sec:demodesign} presents the designed framework of BiTSA. Section \ref{sec:expsetup} details the back-end model training and setup and the model evaluation on downstream tasks. Section \ref{sec:discussion} discusses the strengths and limitations, future directions, and social impact. Finally, the conclusion is provided in Section \ref{sec:conclusion}.

\section{BiTSA Framework Design}
\label{sec:demodesign}

\begin{figure*}[!htb]
    \centering
    \includegraphics[width=0.9\linewidth]{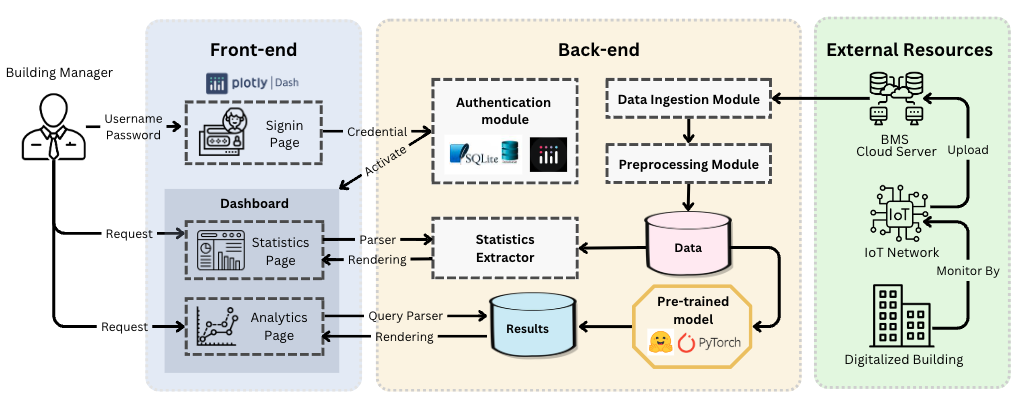}
    \caption{This diagram presents the architecture of BiTSA. The front-end user interface will facilitate user interaction with the system by managing login credentials and handling requests. The back-end will be integrated with the Building Management System to access and process data, execute core functions, and respond to user inputs. Computation results from the back-end will be sent back to the UI for rendering and visualization, ensuring a seamless user experience.}
    \label{fig:dataflow}
\end{figure*}

Figure \ref{fig:dataflow} outlines the pipeline architecture for the BiTSA tool. The system has several core components across front-end, back-end, and connections to external resources, each serving distinct functions to support building management and analytics.

\subsection{Front-end UI Interface}
\begin{figure}
    \centering
    \includegraphics[width=\linewidth]{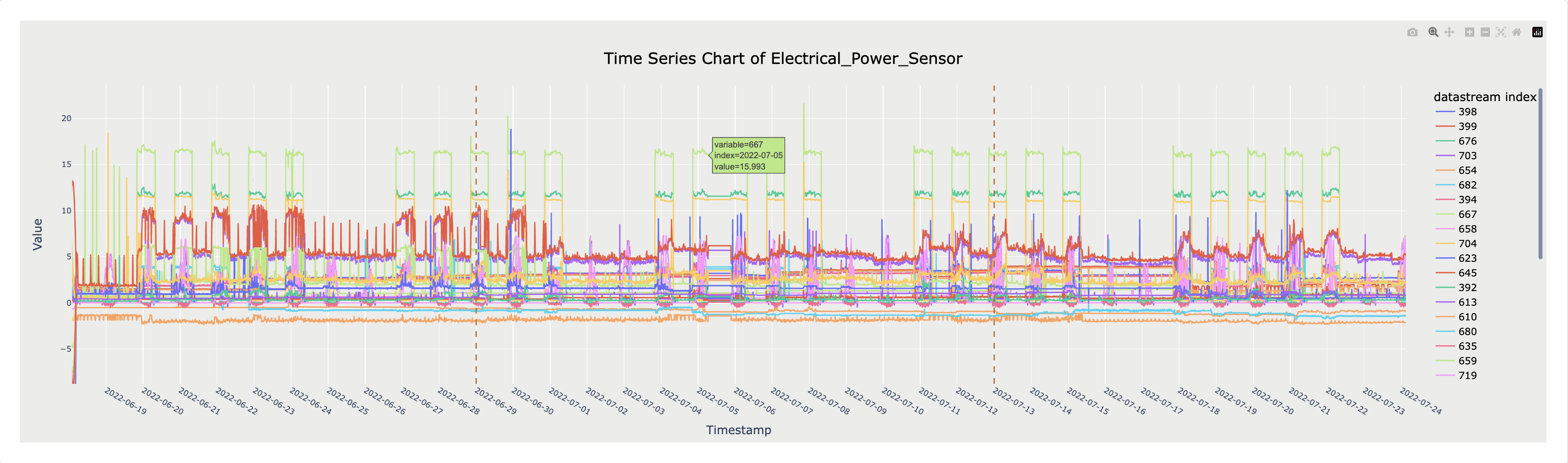}
    \caption{A screenshot of BiTSA displaying interactive time series panel.}
    \label{fig:interactive time series panell}
\end{figure}

\begin{figure}[!htb]
    \centering
    \includegraphics[width=\linewidth]{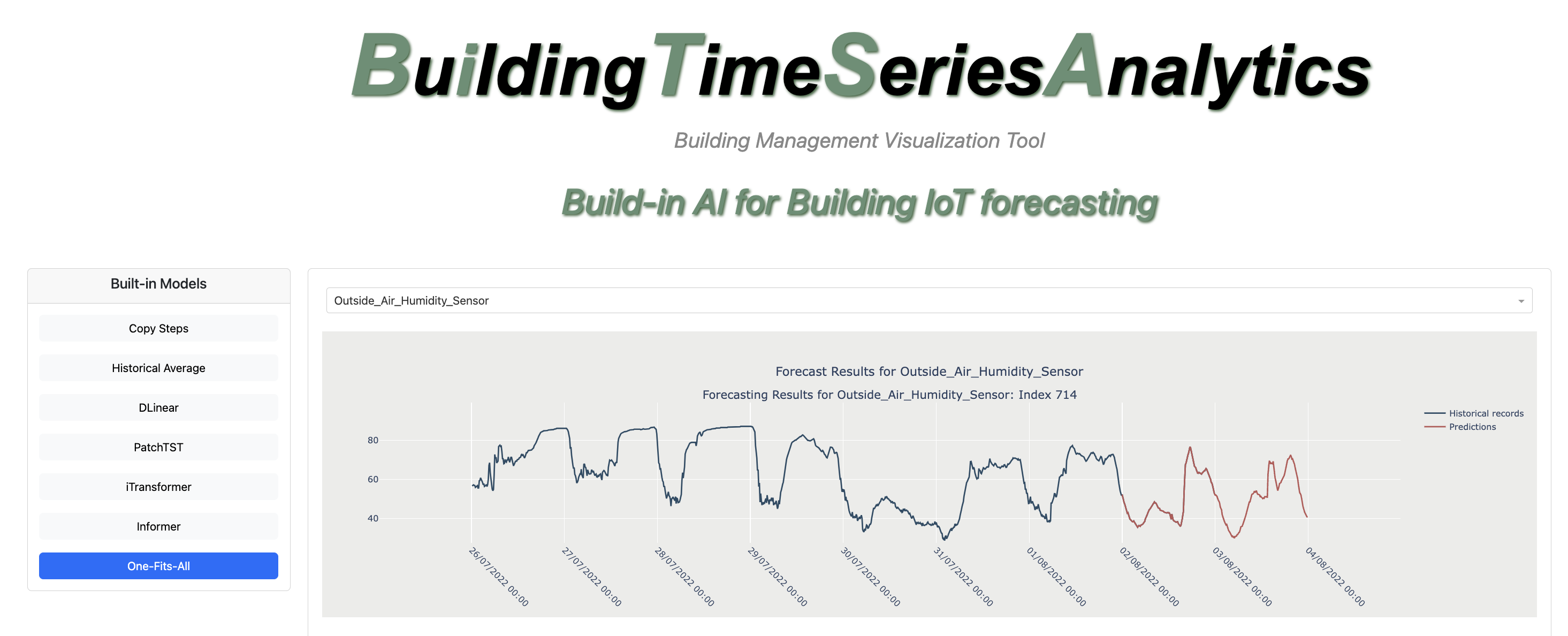}
    \caption{A screenshot of the Analytics page, displaying the predictions of \texttt{Outside\_Air\_Humidity\_Sensor} generated by One-Fits-All. Building manager can filter the sensor through the dropdown to fetch the forecasting results of their interested sensor. }
    \label{fig:analytics page}
\end{figure}

The UI interface will enable building managers to access and interact with visualized forecasting results through a user-friendly dashboard. By handling rendering directly in the browser, the system will deliver a responsive and dynamic experience, allowing users to explore graphs, charts, and other visual elements without the need to refresh the page or send additional server requests while logged in. This approach will provide immediate feedback and support interactive features such as zooming, filtering, and parameter adjustments, as shown in Figure \ref{fig:interactive time series panell}-\ref{fig:analytics page} making it easier for non-technical users to interpret complex forecasting data. The visualizations will offer actionable insights, empowering building managers to make timely and informed decisions regarding energy usage.

\subsection{Back-end Service}
\textbf{Authentication Module}
This module will be responsible for validating users’ credentials. Once logged in successfully, users will gain access to the dashboard and will be able to activate key features to retrieve analytical results.

\textbf{Data Ingestion Module}
BiTSA is designed to integrate with cloud-based databases provided by the user-specific Building Management System (BMS). This module will connect to the Application Programming Interface (API) of the BMS cloud platform and will format the incoming data streams to ensure compatibility for further processing.

\textbf{Preprocessing Module}
The preprocessing module will be responsible for transforming raw IoT-generated data into analytics-ready data, including procedures like granularity alignment, missing values handling, outliers removal, normalization, and feature selection.

\textbf{Statistics Extractor}
This module will retrieve statistical summaries and trends from the analytics-ready data, offering insights into both current and historical building performance. The extracted statistics will be displayed on the dashboard's Statistics Page for access.

\textbf{Pre-trained Models}
The system will support a range of models tailored for various downstream tasks, offering flexibility based on user requirements and application scenarios, as detailed in Section \ref{sec:models}. These models will be pre-trained offline using analytics-ready data and saved within the system, allowing for efficient retrieval and enabling agile deployment.

\textbf{Results Database}
Analytical data processed by the pre-trained models and statistical extractor will be 
stored in a Results database. The results are expected to be updated by each regular period, enabling the analytics and visualizing queries from the user's dashboard.


\subsection{External Resources}
The external resources refer to data collection processes beyond the scope of the tool. Building activities within digital infrastructure will be monitored through an IoT network, with meter readings uploaded to the Building Management Systems (BMS) and stored in cloud-based databases. The connection between BiTSA and the database APIs (i.e, Data Clearing House \footnote{Data Clearing House \url{https://research.csiro.au/dch/}}, Senaps\footnote{Senaps \url{https://products.csiro.au/senaps/}}) will streamline data conversion and ensure continuous updates to support reliable and up-to-date building analytics.

\section{BiTSA Experimental Setup and Performance Evaluation}
\label{sec:expsetup}
Our system will define strict requirements for BMS data formatting, ensuring that IoT-generated data complies with the Brick Schema protocol. Table \ref{tab_stats*} provides an example of existing building IoT datasets that adhere to this standard. 
\begin{table}[htbp]
\centering
\caption{
\textbf{Summary statistics} of the building IoT datasets.
}
\label{tab_stats*}

\begin{tabular}{c||r|rrr}
&
\textbf{BLDG} &
\textbf{BTS-A} &
\textbf{BTS-B} &
\textbf{BTS-C} \\
\toprule

\texttt{Timeseries}			& 337 & 8349 & 851 &  5347 \\
\texttt{Unique Classes}		& 11  & 126  &  57 &  159  \\
\texttt{Country} & USA & \multicolumn{3}{c}{Australia}  \\
\texttt{City} & Berkeley & \multicolumn{3}{c}{Undisclosed}  \\
\texttt{Start Date} & Jan-2018 & Jan-2021 & Jan-2021 & Jun-2021 \\
\texttt{End Date} & Dec-2021 & Dec-2023 & Dec-2023 & Jan-2024 \\
\texttt{Duration} (Days)  & 1094 & 1094 & 1094 & 939 \\
 \bottomrule
\end{tabular}
\end{table}
Since the system is designed for integration with BMS, addressing privacy concerns is critical. To demonstrate functionality, we will employ the BTS-B dataset~\cite{prabowo2024bts} and the BLDG dataset~\cite{luo2022three} (BTS-A/C~\cite{prabowo2024bts} are not yet public) to simulate the respective BMS-control buildings, evaluating the built-in model performance and execute automated pipeline seamlessly.

\subsection{Data Preprocessing}  
The demonstration will process 2-month historical records for each registered building to support model pre-training. Each time series will be resampled to 10-minute intervals, where any missing values will be imputed by second-order polynomial interpolation.

\subsection{Build-in Models} 
\label{sec:models}
Currently, the following models are supported by BiTSA and built into the system: 
DLinear\cite{zeng2023transformers}, PatchTST\cite{nie2022time}, Informer\cite{zhou2021informer}, iTransformer\cite{liu2023itransformer}, and One-Fits-All\cite{zhou2023one} with GPT-2 backbone.
\subsection{Model Training}
Adam optimization with a 1e-3 learning rate guides the learning process. All the models are respectively fed with 1-day historical observations and forecasts on multiple time steps, where $S=144$ and $H\in\{12, 48, 96, 144, 432, 1008\}$, indicating 2-hour, 8-hour, 16-hour, 1-day, 3-day, and 1-week ahead forecasting. The average scores across all the forecasting horizons are computed. The weights are updated using the Adam optimizer with a model-specific learning rate. The number of the training epochs is 100, where EarlyStop with a patience number equal to 10 is employed. The batch size of each model is adjusted to fit the GPU utilization rate. All experiments are completed on V100 or H100 GPUs. 

Mean absolute error (MAE), mean square error (MSE), and symmetric mean absolute percentage error (SMAPE) are employed to evaluate the model performance.

\subsection{Built-in Model Performance}

\begin{table}[!htb]
    \centering
    \begin{tabular}{l|ccc}
    \toprule
         \textbf{Method} & \textbf{MAE} & \textbf{MSE} & \textbf{SMAPE}  \\
         \midrule
         DLinear~\cite{zeng2023transformers} & $0.552\pm0.079$ & \textcolor{red}{$0.669\pm0.111$} & $50.039\pm7.863$ \\
         PatchTST~\cite{nie2022time} & \textcolor{red}{$0.551\pm0.090$} & $0.704\pm0.152$ & $47.463\pm6.370$\\
         Informer~\cite{zhou2021informer} & $0.596\pm0.107$ & $0.752\pm0.163$ & $53.670\pm11.775$ \\
         iTransformer~\cite{liu2023itransformer} & $0.567\pm0.068$ & $0.730\pm0.132$ & $47.577\pm4.358$ \\
         One-Fits-All~\cite{zhou2023one} & $0.552\pm0.089$ & $0.696\pm0.151$ & \textcolor{red}{$44.873\pm6.094$}\\
         \bottomrule
    \end{tabular}
    \vspace{0.2pt}
    \caption{Average results (mean$\pm$std) on $\{$12, 48, 96, 144, 432, 1008$\}$ time-step forecasting on BTS-B. The best score is in \textcolor{red}{red}.}
    \label{tab:bts_res}
\end{table}

\begin{table}[!htb]
    \centering
    \begin{tabular}{l|ccc}
    \toprule
         \textbf{Method} & \textbf{MAE} & \textbf{MSE} & \textbf{SMAPE}  \\
         \midrule
         DLinear~\cite{zeng2023transformers} & \textcolor{red}{$0.547\pm0.132$} & \textcolor{red}{$0.603\pm0.201$} & $48.572\pm11.267$ \\
         PatchTST~\cite{nie2022time} & $0.566\pm0.169$ & $0.711\pm0.332$ & $45.872\pm10.088$ \\
         Informer~\cite{zhou2021informer} & $0.604\pm0.159$ & $0.721\pm0.281$ & $52.833\pm12.943$ \\
         iTransformer~\cite{liu2023itransformer} & $0.582\pm0.177$ & $0.760\pm0.361$ & $45.753\pm10.516$ \\
         One-Fits-All~\cite{zhou2023one} & $0.551\pm0.172$ & $0.710\pm0.351$ & \textcolor{red}{$43.165\pm9.643$} \\
         \bottomrule
    \end{tabular}
    \vspace{0.2pt}
    \caption{Average results (mean$\pm$std) on multiple time-step ahead forecasting on BLDG59. The best performance is in \textcolor{red}{red}.}
    \label{tab:bldg_res}
\end{table}
Tables \ref{tab:bts_res}-\ref{tab:bldg_res} summarize the average performance of built-in models in multi-step forecasting on BTS-B and BLDG59. The results highlight a distinct difference between channel-independent models, such as DLinear, PatchTST, and One-Fits-All, and channel-dependent models, such as Informer and iTransformer. Channel-independent models consistently outperform their channel-dependent counterparts, particularly in capturing time-series patterns within the IoT-generated data. DLinear demonstrates the most robust performance, securing 3 out of the 6 metrics in both datasets. This suggests deploying independent forecasting mechanisms can be more effective in analyzing heterogeneous building IoT-generated data, potentially due to their ability to handle isolated sensor readings without over-complicating the interrelationships between channels. One-Fits-All excels in relative accuracy between predictions and ground truths, indicating its strength in maintaining proportionate accuracy, making it particularly suitable for scenarios where percentage errors are more relevant than absolute differences. A more detailed result is illustrated in~\cite{lin2024exploring}.

In practical terms, building managers can select forecasting models based on their specific operational needs. DLinear would be most suitable in situations where accurate predictions in absolute units are essential. For example, when forecasting energy consumption or cost for budgeting purposes, DLinear’s ability to minimize overall error makes it ideal for tasks where precision in actual quantities is required, which is especially useful for resource allocation, demand planning, or detecting anomalies in energy usage. One-Fits-All is optimal for situations where the relative error or percentage-based improvements are the primary focus. This model could be especially useful for monitoring energy efficiency, benchmarking performance across various community buildings, and tracking sustainability initiatives. In these contexts, understanding trends like percentage deviations or cumulative savings over time is often more informative than focusing solely on precise numerical values.
\subsection{Strengths and Limitations}
\label{sec:discussion}
BiTSA provides significant advantages to building managers by streamlining the analysis of complex energy data and enabling swift, data-driven decisions. Its integration of advanced forecasting models with an intuitive visual interface empowers managers to make proactive choices, optimizing energy consumption and promoting sustainable building practices. BiTSA's capacity to transform abstracted meter readings into actionable insights supports the achievement of energy efficiency targets and facilitates progress toward net-zero goals. By showcasing the power of AI in energy management, BiTSA equips managers with the resources needed for smarter, more sustainable decision-making. This makes it an essential tool for enhancing operational efficiency while advancing sustainability in building management.
One of the limitations of BiTSA is its reliance on offline learning, which often leads to lag in adapting to new data and evolving conditions. This lag can result in delayed responses to changes in building operations or energy patterns, potentially reducing the effectiveness of real-time decision-making. Additionally, BiTSA may lack the ability to provide real-time insights, which are crucial for dynamic environments where immediate adjustments are necessary for optimizing energy efficiency and performance.

\section{Conclusion}
\label{sec:conclusion}
In this paper, we describe the design of BiTSA, an interactive visualization tool equipped with time-series pre-trained models to address challenges associated with the complexity of IoT-generated building data. By bridging the gap between research and real-world application, BiTSA will enable building managers to interpret complex energy data efficiently and to take immediate, data-driven actions based on real-time insights. The integration of advanced forecasting models with an intuitive visual interface will facilitate proactive decision-making, optimize energy consumption, and promote sustainable building management practices. The tool’s capability to transform vast amounts of data into actionable insights will support energy efficiency goals and advance the achievement of net-zero objectives, demonstrating the transformative potential of AI technologies in energy management. Through BiTSA, building managers will be empowered with the tools required to make smarter, more sustainable energy decisions.

\section*{Acknowledgment}
This research is funded by the NSW Government through CSIRO’s NSW Digital Infrastructure Energy Flexibility (DIEF) project as part of the Net Zero Plan Stage 1: 2020-2030, and by the Reliable Affordable Clean Energy for 2030 (RACE for 2030) Cooperative Research Centre.

We also acknowledge the National Computational Infrastructure (NCI Australia) for providing the computational resources necessary to support this work.
\bibliographystyle{IEEEtran}
\bibliography{ref}  


\end{document}